\documentstyle [12pt,amsfonts] {article}
\input epsf
\newcommand{\beq}{\begin{equation}}
\newcommand{\eeq}{\end{equation}}
\newcommand{\beqs}{\begin{eqnarray}}
\newcommand{\eeqs}{\end{eqnarray}}

\newcommand{\gtwid}{\mathrel{\raise.3ex\hbox{$>$\kern-.75em\lower1ex
\hbox{$\sim$}}}}
\newcommand{\ltwid}{\mathrel{\raise.3ex\hbox{$<$\kern-.75em\lower1ex
\hbox{$\sim$}}}}
\newcommand{\bi}{\bibitem}

\catcode`@=11 \@addtoreset{equation}{section}
\@addtoreset{equation}{subsection}
\def\theequation{\ifnum\value{section}=0 \arabic{equation}\ignorespaces
\else \ifnum\value{section}=-1 A.\arabic{equation}\ignorespaces
\else \ifnum\value{subsection}=0
\thesection.\arabic{equation}\ignorespaces \else
\thesection.\arabic{subsection}.\arabic{equation}\ignorespaces
                           \fi
                      \fi
                 \fi}
\catcode`@=12

\begin{document}

\begin{titlepage}

\begin{center}

{\Large \bf Some Exact Formulas on Long-Range Correlation
Functions of the Rectangular Ising Lattice}

\vspace{1.1cm} {\large Shu-Chiuan Chang${}$\footnote{email:
chang@rs.kagu.tus.ac.jp} and Masuo Suzuki${}$\footnote{email:
msuzuki@rs.kagu.tus.ac.jp}}\\
\vspace{18pt}
Department of Applied Physics \\
Faculty of Science \\
Tokyo University of Science \\
Tokyo 162-8601, Japan \\
\end{center}

\vskip 0.6 cm

\begin{abstract}

We study long-range correlation functions of the rectangular Ising
lattice with cyclic boundary conditions. Specifically, we consider
the situation in which two spins are on the same column, and at
least one spin is on or near free boundaries. The low-temperature
series expansions of the correlation functions are presented when
the spin-spin couplings are the same in both directions. The exact
correlation functions can be obtained by D log Pad{\' e} for the
cases with simple algebraic resultant expressions. The present
results show that as the two spins are infinitely far from each
other, the correlation function is equal to the product of the row
magnetizations of the corresponding spins as expected. In terms of
low-temperature series expansions, the approach of this $m$-th row
correlation function to the bulk correlation function for
increasing $m$ can be understood from the observation that the
dominant terms of their series expansions are the same
successively in the above two correlation functions. The number of
these dominant terms increases monotonically as $m$ increases.

\end{abstract}


\end{titlepage}

\section{Introduction}

The Ising model is one of the simplest models in the study of
critical phenomena, and has been investigated by many authors
since Onsager's celebrated work on the free energy of the
rectangular lattice without magnetic field \cite{onsager44},
followed by Yang's derivation of the spontaneous magnetization
\cite{yang52} which gave the critical exponent $\beta = 1/8$. The
correlation function $<\sigma _{0,0} \sigma _{m,n}>$ of the
rectangular lattice has also been well studied, and can be
expressed in terms of Toeplitz determinantal form with dimension
$2m+2n$ \cite{montroll63,mccoybook,schults64} and a "generalized
Wronskian" form with dimension $m$ \cite{yamada86}. The most
efficient way to calculate the bulk correlation functions is given
in Refs. \cite{orrick01,auyang02} and the references therein.
McCoy and Wu considered the rectangular lattice with cyclic
boundary conditions, and only the spins on a boundary row interact
with a magnetic field \cite{mccoy67}. In this case, the critical
exponent of the boundary spontaneous magnetization was found to be
1/2, and the magnetization in the $m$-th row was expressed by a
determinant with dimension $m$. However, the approach of the
boundary spontaneous magnetization to the bulk result was
unsolved. As the square of the spontaneous magnetization is equal
to the long-range correlation function, this problem is equivalent
to the approach of the boundary correlation function to the bulk
correlation function. It is the purpose of this paper to study
this approach by the long-range correlation functions with at
least one spin on or near the free boundaries. In section 2, we
give the known and expected results of the correlation functions
together with the basic definition. The calculation and
expressions of the correlation functions between two spins on a
column will be sketched in section 3. Since the correlation
functions do not have a simple closed form in general, we content
ourselves with the expression of the correlation functions in
terms of low-temperature series expansions, and indicate the
approach of the boundary correlation function to the bulk
correlation function in section 4. Conclusions will be given in
Section 5.

\section{Rectangular Ising lattice and the correlation functions}

We mainly consider here a rectangular lattice with cyclic boundary
conditions imposed in the horizontal direction, and free boundary
conditions in the vertical direction without magnetic field. The
numbers of rows and columns are denoted by ${\cal M}$ and ${\cal
N}$, respectively. The Hamiltonian of the system is
\beq {\cal E} = -J_1 \sum _{j=1}^{\cal M} \sum _{k=1}^{\cal N}
\sigma _{j,k}\sigma _{j,k+1} -J_2 \sum _{j=1}^{{\cal M}-1} \sum
_{k=1}^{\cal N} \sigma _{j,k}\sigma _{j+1,k} \label{hamiltonian}
\eeq
where $J_1$ and $J_2$ are the horizontal and vertical spin-spin
interactions between nearest-neighbor spins, and $\sigma _{j,{\cal
N}+1} = \sigma _{j,1}$. Each $\sigma$ assumes the value $+1$ or
$-1$.

In the two-dimensional limit, i.e., ${\cal M} \to \infty$ and
${\cal N} \to \infty$, the solution of the spontaneous
magnetization is well-known \cite{yang52,montroll63,onsager49},
namely the bulk correlation function $<\sigma _{m,1} \sigma
_{m,n}>$ between two spins on the same row in the limits $n \to
\infty$, ${\cal N} - n \to \infty$ and $m \to \infty$, ${\cal M} -
m \to \infty$ is
\beqs <\sigma _{m,1} \sigma _{m,n}> = M_{\scriptsize
\mbox{s,bulk}} ^2 & = & \left [ 1 -
\frac{(1-z_1^2)^2(1-z_2^2)^2}{16z_1^2z_2^2} \right ] ^{1/4} \cr\cr
& = & \left [ 1 - \frac{1}{\sinh ^2 (2K_1) \sinh ^2 (2K_2)} \right
] ^{1/4} \label{correlationrowbulk} \eeqs
where $K_1=\beta J_1, K_2=\beta J_2$ and $z_1=\tanh K_1, z_2=\tanh
K_2$.

The boundary correlation function $<\sigma _{1,1} \sigma _{1,n}>$
was considered in \cite{mccoy67}. In the limit $n \to \infty$ and
${\cal N} - n \to \infty$,
\beqs <\sigma _{1,1} \sigma _{1,n}> = M_{\scriptsize \mbox{s,1}}
^2 & = & \frac{z_2^2(1+z_1)^2 - (1-z_1)^2}{4z_1 z_2^2} \cr\cr & =
& \frac{\cosh (2K_2) - \coth (2K_1)}{\cosh (2K_2) - 1} \ .
\label{correlationrowboundary} \eeqs
The second-row correlation function $<\sigma _{2,1} \sigma
_{2,n}>$ in the limit $n \to \infty$, ${\cal N} - n \to \infty$ is
\cite{mccoy67}
\beqs & & <\sigma _{2,1} \sigma _{2,n}> = M_{\scriptsize
\mbox{s,2}} ^2 \cr\cr & = & \frac{z_2^2(1+z_1)^2 - (1-z_1)^2}{4z_1
z_2^4} \left [ 1 - \frac{z_1(1-z_2^2)}{z_2(1-z_1^2)}
\frac{1}{2\pi} \int _{-\pi}^{\pi} d\theta \alpha
^{-1}(e^{i\theta}-1)(e^{-i\theta}-1) \right ]^2 \cr & &
\label{correlationrow2} \eeqs
where $\alpha$ is the larger root in magnitude of the equation
\beq (1+z_1^2)(1+z_2^2) - z_1(1-z_2^2)(e^{i\theta} + e^{-i\theta})
- z_2(1-z_1^2)(\alpha + \alpha ^{-1}) = 0 \ . \eeq

Let us consider the long-range correlation function between two
spins on the same column. In the limit that the distance between
these spins is infinite, the correlation function should be the
product of the magnetization of each spin. Therefore, the bulk
correlation function $<\sigma _{m^\prime,1} \sigma _{m,1}>$ with
$m^\prime < m$ in the limits $m-m^\prime \to \infty$, $m^\prime
\to \infty$, ${\cal M} - m \to \infty$ should be $<\sigma
_{m^\prime,1} \sigma _{m,1}> = M_{\scriptsize \mbox{s,bulk}} ^2$.
Similarly, we expect the following results: the boundary-boundary
correlation function $\lim _{{\cal M} \to \infty} <\sigma _{1,1}
\sigma _{{\cal M},1}> = M_{\scriptsize \mbox{s,1}} ^2$, the second
row-second row correlation function $\lim _{{\cal M} \to \infty}
<\sigma _{2,1} \sigma _{{\cal M} -1,1}> = M_{\scriptsize
\mbox{s,2}} ^2$, and the boundary-second row correlation function
$\lim _{{\cal M} \to \infty} <\sigma _{2,1} \sigma _{{\cal M},1}>
= M_{\scriptsize \mbox{s,1}} M_{\scriptsize \mbox{s,2}}$, etc. If
one of the spins is far from the free boundaries, say on the
middle row, then we expect that the boundary-middle correlation
function $\lim _{{\cal M} \to \infty} <\sigma _{1,1} \sigma
_{[{\cal M}/2],1}> = M_{\scriptsize \mbox{s,1}} M_{\scriptsize
\mbox{s,bulk}}$ and the second row-middle correlation function
$\lim _{{\cal M} \to \infty} <\sigma _{2,1} \sigma _{[{\cal
M}/2],1}> = M_{\scriptsize \mbox{s,2}} M_{\scriptsize
\mbox{s,bulk}}$, etc., where $M_{\scriptsize \mbox{s,bulk}},
M_{\scriptsize \mbox{s,1}}, M_{\scriptsize \mbox{s,2}}$ are given
in eqns. (\ref{correlationrowbulk}) to (\ref{correlationrow2}) and
$[g]$ denotes the integral part of $g$. Whether ${\cal M}$ is even
or odd is not important here since we take the limit ${\cal M} \to
\infty$.

\section{Calculation for the correlation function between two spins on the same column}

We calculate correlation functions between two spins on the same
column by the standard Pfaffian approach. Since this method is
available in the literature
\cite{montroll63,mccoybook,mccoy67,kasteleyn63} for decades, the
derivation of our results should be brief. As is well-known, the
rectangular Ising lattice is equivalent to the dimer problem on
the "bathroom tile" lattice. The square of the partition function
is given by
\beq Z^2 = (2\cosh K_1)^{2{\cal NM}} (\cosh K_2)^{2{\cal N(M}-1)}
\det {\cal A} \label{partitionnsqrt} \eeq
where ${\cal A}$ is an antisymmetrical matrix, and some of its
elements in the limit ${\cal N} \to \infty$ will be given
explicitly in eqns. (\ref{ainverse1}) to (\ref{ainverse4}). The
square of the correlation function between $\sigma _{m^\prime,1}$
and $\sigma _{m,1}$ with $m^\prime < m$ can be shown to be
\beq <\sigma _{m^\prime,1} \sigma _{m,1}>^2 = z_2^{2(m-m^\prime)}
\det (\bar{\delta}^{-1} + \bar{\cal A}^{-1}) \det \bar{\delta}
\eeq
where in the standard notation
\oddsidemargin=-0.6in
\beq \bar{\delta} = \!\!\!\!
\begin{array}{cc} & \begin{array}{cccccccc}
(m^\prime,1)_U & (m^\prime+1,1)_U & \cdots &
 (m-1,1)_U & (m^\prime+1,1)_D & (m^\prime+2,1)_D & \cdots & (m,1)_D
\end{array} \\
\begin{array}{c} (m^\prime,1)_U \!\!\!\!\!\!\!\! \\ (m^\prime+1,1)_U \!\!\!\!\!\!\!\! \\
\vdots \!\!\!\!\!\!\!\! \\ (m-1,1)_U \!\!\!\!\!\!\!\! \\ (m^\prime+1,1)_D \!\!\!\!\!\!\!\! \\
(m^\prime+2,1)_D \!\!\!\!\!\!\!\! \\ \vdots \!\!\!\!\!\!\!\! \\
(m,1)_D \!\!\!\!\!\!\!\!
\end{array} & \left [
\begin{array}{cccccccc} 0 & 0 & \cdots & 0 & \ \ z_2^{-1}-z_2 \ \
& 0 & \cdots & 0
\\ 0 & 0 & \cdots & 0 & 0 & \ \ z_2^{-1}-z_2 \ \ \, & \cdots & 0
\\ \vdots & \vdots & \ddots & \vdots & \vdots & \vdots & \ddots & \vdots
\\ 0 & 0 & \cdots & 0 & 0 & 0 & \cdots & z_2^{-1}-z_2
\\ -z_2^{-1}+z_2 & 0 & \cdots & 0 & 0 & 0 & \cdots & 0
\\ 0 & -z_2^{-1}+z_2 \ & \cdots & 0 & 0 & 0 & \cdots & 0
\\ \vdots & \vdots & \ddots & \vdots & \vdots & \vdots & \ddots &
\vdots \\ 0 & 0 & \cdots & -z_2^{-1}+z_2 \ \ \ & 0 & 0 & \cdots &
0 \end{array} \right ] \end{array} \\ \eeq
so that $\det \bar{\delta} = (z_2 ^{-1} - z_2)^{2(m-m^\prime)}$,
and $\bar{\cal A}^{-1}$ is the submatrix of ${\cal A}^{-1}$ in the
subspace $\{ (m^\prime ,1)_U, ..., (m,1)_D\}$. Explicitly, we have
\beqs & & <\sigma _{m^\prime,1} \sigma _{m,1}>^2 =
(1-z_2^2)^{2(m-m^\prime)} \times \cr\cr & & \left |
\begin{array}{cccccc} (m^\prime;m^\prime)_{UU} & \cdots & (m^\prime;m-1)_{UU} &
(m^\prime;m^\prime+1)_{UD} - \bar{z}_2 & \cdots & (m^\prime;m)_{UD} \\
(m^\prime+1;m^\prime)_{UU} & \cdots & (m^\prime+1;m-1)_{UU} &
(m^\prime+1;m^\prime+1)_{UD} & \cdots & (m^\prime+1;m)_{UD} \\
\vdots & \ddots & \vdots & \vdots & \ddots & \vdots \\
(m-1;m^\prime)_{UU} & \cdots & (m-1;m-1)_{UU} &
(m-1;m^\prime+1)_{UD} & \cdots & (m-1;m)_{UD} - \bar{z}_2 \\
(m^\prime+1;m^\prime)_{DU} + \bar{z}_2 & \cdots &
(m^\prime+1;m-1)_{DU}
& (m^\prime+1;m^\prime+1)_{DD} & \cdots & (m^\prime+1;m)_{DD} \\
(m^\prime+2;m^\prime)_{DU} & \cdots & (m^\prime+2;m-1)_{DU} &
(m^\prime+2;m^\prime+1)_{DD} & \vdots & \vdots \\
\vdots & \ddots & \vdots & \vdots & \ddots & \vdots \\
(m;m^\prime)_{DU} & \cdots & (m;m-1)_{DU} + \bar{z}_2 &
(m;m^\prime+1)_{DD} & \cdots & (m;m)_{DD}
\end{array} \right | \cr\cr & & \eeqs
where we use the abbreviations $\bar{z}_2 \equiv
(z_2^{-1}-z_2)^{-1}$ and $(j;j^\prime)_{XX^\prime} \equiv {\cal
A}^{-1}(j,1;j^\prime,1)_{XX^\prime}$ with $X, X^\prime \in \{U, D
\}$. By a procedure similar to that in \cite{mccoy67}, the
elements of the inverse matrix ${\cal A}^{-1}$ can be calculated.
In the limit ${\cal N} \to \infty$, they are
\beqs & & {\cal A}^{-1}(j,k;j^\prime,k^\prime)_{DD} = -{\cal
A}^{-1}(j^\prime,k^\prime;j,k)_{DD} \cr\cr & = & -\frac{1}{2\pi}
\int _{-\pi} ^{\pi} d\theta e^{i\theta(k-k^\prime)} \frac{\alpha
^{j^\prime-j}(e^{i\theta}-e^{-i\theta})z_1}{(\alpha
^{-1}-\alpha)z_2(1-z_1^2)} \frac{[1+\alpha ^{-2(j^\prime -1)}
(\frac{c^\prime}{c})^2] [1-\alpha ^{-2({\cal M}-j+1)} ]}{1+\alpha
^{-2{\cal M}}(\frac{c^\prime}{c})^2} \label{ainverse1} \eeqs
\beqs & & {\cal A}^{-1}(j,k;j^\prime,k^\prime)_{UU} = -{\cal
A}^{-1}(j^\prime,k^\prime;j,k)_{UU} \cr\cr & = & \frac{1}{2\pi}
\int _{-\pi} ^{\pi} d\theta e^{i\theta(k-k^\prime)} \frac{\alpha
^{j^\prime-j}(e^{i\theta}-e^{-i\theta})z_1}{(\alpha
^{-1}-\alpha)z_2(1-z_1^2)} \frac{[1-\alpha ^{-2j^\prime}]
[1+\alpha ^{-2({\cal M}-j)}(\frac{c^\prime}{c})^2 ]}{1+\alpha
^{-2{\cal M}}(\frac{c^\prime}{c})^2} \label{ainverse2} \eeqs
\beqs & & {\cal A}^{-1}(j,k;j^\prime,k^\prime)_{UD} = -{\cal
A}^{-1}(j^\prime,k^\prime;j,k)_{DU} \cr\cr & = & -\frac{1}{2\pi}
\int _{-\pi} ^{\pi} d\theta e^{i\theta(k-k^\prime)} \frac{\alpha
^{j^\prime-j}(1-z_1^2 -
|1+z_1e^{i\theta}|^2\frac{z_2}{\alpha})}{(\alpha
^{-1}-\alpha)z_2(1-z_1^2)} \frac{[1+\alpha ^{-2(j^\prime -1)}
(\frac{c^\prime}{c})^2] [1+\alpha ^{-2({\cal
M}-j)}(\frac{c^\prime}{c})^2 ]}{1+\alpha ^{-2{\cal
M}}(\frac{c^\prime}{c})^2} \cr\cr & & \label{ainverse3} \eeqs
\beqs & & {\cal A}^{-1}(j,k;j^\prime,k^\prime)_{DU} = -{\cal
A}^{-1}(j^\prime,k^\prime;j,k)_{UD} \cr\cr & = & \frac{1}{2\pi}
\int _{-\pi} ^{\pi} d\theta e^{i\theta(k-k^\prime)} \frac{\alpha
^{j^\prime-j}(1-z_1^2 - |1+z_1e^{i\theta}|^2z_2\alpha)}{(\alpha
^{-1}-\alpha)z_2(1-z_1^2)} \frac{[1-\alpha ^{-2j^\prime}]
[1-\alpha ^{-2({\cal M}-j+1)} ]}{1+\alpha ^{-2{\cal
M}}(\frac{c^\prime}{c})^2} \label{ainverse4} \eeqs
\oddsidemargin=-0.2in
where $1 \le j^\prime \le j$ for the first three functions, and $1
\le j^\prime < j$ for the last one. $c$ and $c^\prime$ are defined
as
\beq c = \sqrt{\frac{1}{2}\left ( 1 + \frac{z_2^2+a^2-b^2}{\lambda
^\prime - \lambda} \right )} \ , \qquad c^\prime = \mbox{sgn}
(iaz_2) \sqrt{\frac{1}{2}\left ( 1 - \frac{z_2^2+a^2-b^2}{\lambda
^\prime - \lambda} \right )} \label{cdef} \eeq
where $a = 2iz_1 \sin \theta |1+z_1e^{i\theta}|^{-2}$, $b =
(1-z_1^2)|1+z_1e^{i\theta}|^{-2}$, and $\lambda =
|1+z_1e^{i\theta}|^{-2}z_2 (1-z_1^2)\alpha$, $\lambda ^\prime =
|1+z_1e^{i\theta}|^{-2}z_2 (1-z_1^2)\alpha ^{-1}$.

Because ${\cal A}^{-1}(j,k;j^\prime,k)_{DD} = {\cal
A}^{-1}(j,k;j^\prime,k)_{UU} = 0$ and the sign of $<\sigma
_{m^\prime,1} \sigma _{m,1}>$ should be chosen so that $<\sigma
_{m^\prime,1} \sigma _{m,1}> \to 1$ as $T \to 0$, we obtain the
following general expression of the correlation function between
two spins on the same column,
\beqs & & <\sigma _{m^\prime,1} \sigma _{m,1}> = \cr\cr & & \left
|
\begin{array}{cccc} z_2-\tilde{z}_2{\cal A}^{-1}(m^\prime;m^\prime+1)_{UD}
& -\tilde{z}_2{\cal A}^{-1}(m^\prime;m^\prime+2)_{UD} & \cdots &
-\tilde{z}_2{\cal A}^{-1}(m^\prime;m)_{UD} \\
-\tilde{z}_2{\cal A}^{-1}(m^\prime+1;m^\prime+1)_{UD} &
z_2-\tilde{z}_2{\cal A}^{-1}(m^\prime+1;m^\prime+2)_{UD} &
\cdots & -\tilde{z}_2{\cal A}^{-1}(m^\prime+1;m)_{UD} \\
\vdots & \vdots & \ddots & \vdots \\
-\tilde{z}_2{\cal A}^{-1}(m-1;m^\prime+1)_{UD} & -\tilde{z}_2{\cal
A}^{-1}(m-1;m^\prime+2)_{UD} & \cdots & z_2-\tilde{z}_2{\cal
A}^{-1}(m-1;m)_{UD} \end{array} \right| \cr\cr & &
\label{generalcf} \eeqs
where we define
\beqs & & -\tilde{z}_2 {\cal A}^{-1}(j;j^\prime)_{UD} \equiv
-(1-z_2^2){\cal A}^{-1}(j,1;j^\prime,1)_{UD} \cr\cr & = &
\frac{1}{2\pi} \int _{-\pi} ^{\pi} d\theta \frac{\alpha
^{j^\prime-j}(1-z_2^2)}{(\alpha ^{-1}-\alpha)z_2(1-z_1^2)}[1-z_1^2
- |1+z_1e^{i\theta}|^2\frac{z_2}{\alpha}] \frac{[1+\alpha
^{-2(j^\prime -1)} (\frac{c^\prime}{c})^2] [1+\alpha ^{-2({\cal
M}-j)}(\frac{c^\prime}{c})^2 ]}{1+\alpha ^{-2{\cal
M}}(\frac{c^\prime}{c})^2} \cr\cr & & \qquad \qquad \mbox{for}
\quad1 \le j^\prime \le j \label{elementll} \eeqs
\beqs & & -\tilde{z}_2 {\cal A}^{-1}(j^\prime;j)_{UD} \equiv
-(1-z_2^2){\cal A}^{-1}(j^\prime,1;j,1)_{UD} \cr\cr & = &
\frac{1}{2\pi} \int _{-\pi} ^{\pi} d\theta \frac{\alpha
^{j^\prime-j}(1-z_2^2)}{(\alpha ^{-1}-\alpha)z_2(1-z_1^2)}[1-z_1^2
- |1+z_1e^{i\theta}|^2z_2\alpha] \frac{[1-\alpha ^{-2j^\prime}]
[1-\alpha ^{-2({\cal M}-j+1)} ]}{1+\alpha ^{-2{\cal
M}}(\frac{c^\prime}{c})^2} \cr\cr & & \qquad \qquad \mbox{for}
\quad 1 \le j^\prime < j \ . \label{elementur} \eeqs

\section{Correlation functions in terms of low-temperature series expansions}

The correlation function $<\sigma _{m^\prime,1} \sigma _{m,1}>$
given in eqn. (\ref{generalcf}) is expressed by the determinant of
a $(m-m^\prime) \times (m-m^\prime)$ matrix. As we are interested
in long-range correlation functions, the size of the matrix
approaches infinity. Instead of performing the integrations of all
the matrix elements directly, we expand the integrands in terms of
low-temperature expansions. For simplicity, we consider $J_1 = J_2
= J$, and use the low-temperature variable $z=e^{-2\beta J}$. It
is expected that the first few terms of the series for a finite
value of $m-m^\prime$ become the same as in the infinite case and
that number of correct terms would increase as the difference
$m-m^\prime$ increases. If the exact result has such a simple
expression as in eqns. (\ref{correlationrowbulk}) or
(\ref{correlationrowboundary}), it can be obtained by D log
Pad{\'e} analysis \cite{gaunt74dombbook} with the first certain
terms known. We shall check this procedure on the bulk correlation
function first. Then we apply it to other cases in this section.

\subsection{Bulk case}

For the bulk correlation function, we take the limits $m-m^\prime
\to \infty$, $m^\prime \to \infty$, ${\cal M} - m \to \infty$. The
last two limits will not be written out explicitly in this
subsection. In these limits, the last three factors in eqns.
(\ref{elementll}) and (\ref{elementur}) are equal to 1. Using the
notations $f_{j-j^\prime}$ and $f_{j-j^\prime}^\prime$ for the
corresponding matrix elements $(1-z_2^2){\cal
A}^{-1}(j^\prime,1;j,1)_{UD}$ for $1 \le j^\prime < j$ and
$(1-z_2^2){\cal A}^{-1}(j,1;j^\prime,1)_{UD}$ for $1 \le j^\prime
\le j$, respectively, the bulk correlation function is given by
\beq \lim _{m-m^\prime \to \infty} <\sigma _{m^\prime,1} \sigma
_{m,1}> = \lim _{m-m^\prime \to \infty} \left |
\begin{array}{ccccc} z_2-f_1 & -f_2 & -f_3 & \cdots & -f_{m-m^\prime}
\\ -f_0^\prime & z_2-f_1 & -f_2 & \cdots & -f_{m-m^\prime-1}
\\ -f_1^\prime & -f_0^\prime & z_2-f_1 & \ddots & \vdots
\\ \vdots & \vdots & \ddots & \ddots & -f_2
\\ -f_{m-m^\prime-2}^\prime & -f_{m-m^\prime-3}^\prime & \cdots
& -f_0^\prime & z_2-f_1 \end{array} \right | \ . \eeq
This is the traditional Toeplitz determinant, and the matrix
corresponds to the infinite-size center part of the matrix for
$\lim _{{\cal M} \to \infty} <\sigma _{1,1} \sigma _{{\cal M},1}>$
as shown in the next subsection. We consider here the $\ell \times
\ell$ submatrix of the above matrix by keeping the first $\ell$
columns and the first $\ell$ rows where $\ell = m-m^\prime$ is
finite. Then, we define the correlation function $\lim _{{\cal M}
\to \infty} <\sigma _{m^\prime,1} \sigma _{m,1}>_{\ell}$ by the
determinant of that $\ell \times \ell$ submatrix. By expanding the
integrand of each matrix element in the power series of $z$, by
performing the integrations and by computing the determinant, we
obtain
\beqs \lim _{{\cal M} \to \infty} <\sigma _{m^\prime,1} \sigma
_{m,1}>_1 & = & z_2-f_1 =
1-4z^4-12z^6-36z^8-120z^{10}-448z^{12}-1820z^{14}+... \cr\cr \lim
_{{\cal M} \to \infty} <\sigma _{m^\prime,1} \sigma _{m,1}>_2 & =
& 1-4z^4-16z^6-60z^8-224z^{10}-860z^{12}-3472z^{14}+... \cr\cr
\lim _{{\cal M} \to \infty} <\sigma _{m^\prime,1} \sigma _{m,1}>_3
& = & 1-4z^4-16z^6-64z^8-268z^{10}-1152z^{12}-4960z^{14}+...
\cr\cr \lim _{{\cal M} \to \infty} <\sigma _{m^\prime,1} \sigma
_{m,1}>_4 & = &
1-4z^4-16z^6-64z^8-272z^{10}-1224z^{12}-5680z^{14}+... \cr\cr \lim
_{{\cal M} \to \infty} <\sigma _{m^\prime,1} \sigma _{m,1}>_5 & =
& 1-4z^4-16z^6-64z^8-272z^{10}-1228z^{12}-5788z^{14}+... \cr\cr
\lim _{{\cal M} \to \infty} <\sigma _{m^\prime,1} \sigma _{m,1}>_6
& = & 1-4z^4-16z^6-64z^8-272z^{10}-1228z^{12}-5792z^{14}+...
\label{finiteb} \eeqs
Compared with the exact bulk correlation function
\beqs \lim _{m-m^\prime \to \infty} <\sigma _{m^\prime,1} \sigma
_{m,1}> & = & \left [1 - \frac{16z^4}{(1-z^2)^4} \right ]^{1/4}
\cr\cr & = & 1-4z^4-16z^6-64z^8-272z^{10}-1228z^{12}-5792z^{14}
\cr\cr & & -28192z^{16}-140448z^{18}-712276z^{20}+...
\label{bulks} \eeqs
We find that the series of $\lim _{{\cal M} \to \infty} <\sigma
_{m^\prime,1} \sigma _{m,1}>_{\ell}$ is correct up to the
$z^{2\ell +2}$ term. Because the coordination number of
non-boundary spins is 4, the power of $z$ in the series expansion
terms for the bulk correlation function is always even. In this
case, $u=z^2$ is used as the expansion variable traditionally.
Notice that this is no long true for the other cases we consider
in the rest of this section. Since

\beq \frac{d}{du} \ln \Bigl ( \lim _{m-m^\prime \to \infty}
<\sigma _{m^\prime,1} \sigma _{m,1}> \Bigr ) =
\frac{-8u}{(1-u^2)(1-6u+u^2)} \ , \eeq
one only needs terms up to $z^{12}$ to obtain the exact result by
D log pad{\' e} because only 5 unknown parameters in this equation
need to be determined.

\subsection{Boundary-boundary case}

In order to obtain the boundary-boundary correlation function, we
substitute $m^\prime =1$ and $m={\cal M}$ in eqn.
(\ref{generalcf}). We show here explicitly the power series of the
correlation function $<\sigma _{1,1} \sigma _{{\cal M},1}>_{\cal
M}$ which is given by the determinant of a $({\cal M} -1) \times
({\cal M} -1)$ matrix (\ref{generalcf}):
\beqs <\sigma _{1,1} \sigma _{2,1}>_2 & = &
1-4z^3-8z^4-4z^5+48z^6+184z^7+272z^8+... \cr\cr <\sigma _{1,1}
\sigma _{3,1}>_3 & = & 1-4z^3-8z^4-16z^5-24z^6-16z^7+152z^8+...
\cr\cr <\sigma _{1,1} \sigma _{4,1}>_4 & = &
1-4z^3-8z^4-16z^5-24z^6-36z^7-48z^8+... \cr\cr <\sigma _{1,1}
\sigma _{5,1}>_5 & = &
1-4z^3-8z^4-16z^5-24z^6-36z^7-48z^8-64z^9-80z^{10}+... \cr\cr
<\sigma _{1,1} \sigma _{6,1}>_6 & = &
1-4z^3-8z^4-16z^5-24z^6-36z^7-48z^8-64z^9-80z^{10}-100z^{11}
\cr\cr & & -120z^{12}+... \label{finite11} \eeqs
Compared with the following exact result in the limit ${\cal M}
\to \infty$
\beqs \lim _{{\cal M} \to \infty} <\sigma _{1,1} \sigma _{{\cal
M},1}> & = & \frac{(1 - 2z-z^2)(1+z^2)}{(1-z)^3(1+z)} \cr\cr & = &
1-4z^3-8z^4-16z^5-24z^6-36z^7-48z^8-64z^9 \cr\cr & &
-80z^{10}-100z^{11}-120z^{12}-144z^{13}-168z^{14}+... \label{bbs}
\eeqs
we observe that $<\sigma _{1,1} \sigma _{{\cal M},1}>_{\cal M}$ is
correct to the $z^{2{\cal M}}$ term. Now one only needs terms up
to $z^{11}$ to obtain the exact result by D log Pad{\' e} because
\beq \frac{d}{dz} \ln \Bigl ( \lim _{{\cal M} \to \infty} <\sigma
_{1,1} \sigma _{{\cal M},1}> \Bigr ) =
\frac{-8z^2(3+2z+z^2)}{(1-z^4)(1-2z-z^2)} \ . \eeq

\subsection{Second row-second row and higher row cases}

For the second row-second row correlation function, $<\sigma
_{2,1} \sigma _{{\cal M}-1,1}>$, its matrix corresponds to the
matrix for $<\sigma _{1,1} \sigma _{{\cal M},1}>$ with the first
and the last columns and rows removed. Again consider finite
${\cal M}$ so that $<\sigma _{2,1} \sigma _{{\cal M}-1,1}>_{\cal
M}$ is the determinant of the $({\cal M} -3) \times ({\cal M} -3)$
matrix. The cases ${\cal M} = 4,5$ and $6$ are given as
\beqs <\sigma _{2,1} \sigma _{3,1}>_4 & = &
1-4z^4-4z^5-24z^6-52z^7-112z^8+... \cr\cr <\sigma _{2,1} \sigma
_{4,1}>_5 & = &
1-4z^4-4z^5-28z^6-56z^7-156z^8-296z^9-576z^{10}+... \cr\cr <\sigma
_{2,1} \sigma _{5,1}>_6 & = &
1-4z^4-4z^5-28z^6-56z^7-160z^8-300z^9-648z^{10} \cr\cr & &
-1124z^{11}-2076z^{12}+... \eeqs
These series expansions approach the expected result for ${\cal M}
\to \infty$ which is
\beqs \lim _{{\cal M} \to \infty} < \sigma _{2,1} \sigma _{{\cal
M}-1,1}> & = &
1-4z^4-4z^5-28z^6-56z^7-160z^8-300z^9-652z^{10}-1128z^{11} \cr\cr
& & -2188z^{12}-3500z^{13}-6340z^{14}+... \eeqs
Notice that there is no $z^3$ term in contrast with the
boundary-boundary case in eqn. (\ref{bbs}), and the coefficient of
$z^4$ term is the same as the coefficient of $z^4$ term for the
bulk case in eqn. (\ref{bulks}).

In general, the matrix for $<\sigma _{m+1,1} \sigma _{{\cal
M}-m,1}>$ corresponds to the matrix for $<\sigma _{1,1} \sigma
_{{\cal M},1}>$ with the first and the last $m$ columns and rows
removed, and we expect that $\lim _{{\cal M} \to \infty} <\sigma
_{m+1,1} \sigma _{{\cal M}-m,1}>$ is the same as the square of the
$(m+1)$-th row magnetization. However, there may be no simple
expressions for these correlation functions, and these can not be
obtained by D log Pad{\' e} analysis.

As the next example, the matrix for the third row-third row
correlation function is the submatrix of the matrix for $<\sigma
_{1,1} \sigma _{{\cal M},1}>$ with the first and the last two
columns and rows removed. The first two series expansions of
$<\sigma _{3,1} \sigma _{{\cal M}-2,1}>_{\cal M}$ with finite
${\cal M}$ are
\beqs <\sigma _{3,1} \sigma _{4,1}>_6 & = &
1-4z^4-12z^6-4z^7-56z^8-124z^9-432z^{10}+... \cr\cr <\sigma _{3,1}
\sigma _{5,1}>_7 & = &
1-4z^4-16z^6-4z^7-80z^8-128z^9-564z^{10}+... \eeqs
In this case, there is no $z^3$ and $z^5$ terms, and the
coefficients of $z^4$ and $z^6$ terms are the same as the
corresponding coefficients for the bulk case in eqn.
(\ref{bulks}). Therefore, we expect that the number of such
leading terms in the series expansion of $\lim _{{\cal M} \to
\infty} <\sigma _{m+1,1} \sigma _{{\cal M}-m,1}>$ as agree with
the corresponding terms for the bulk case increases monotonically
as $m$ increases.

\subsection{Boundary-second row and boundary-higher row cases}

For the boundary-second row correlation function, $<\sigma _{2,1}
\sigma _{{\cal M},1}>$, its matrix corresponds to the matrix for
$<\sigma _{1,1} \sigma _{{\cal M},1}>$ with the first column and
row removed. For finite ${\cal M}$, the correlation function
$<\sigma _{2,1} \sigma _{{\cal M},1}>_{\cal M}$ is given by the
determinant of the $({\cal M} -2) \times ({\cal M} -2)$ matrix.
The power series expansions for small values of ${\cal M}$ are
given in the following:
\beqs <\sigma _{2,1} \sigma _{3,1}>_3 & = &
1-2z^3-6z^4-8z^5-18z^6-20z^7+56z^8+... \cr\cr <\sigma _{2,1}
\sigma _{4,1}>_4 & = & 1-2z^3-6z^4-10z^5-28z^6-48z^7-104z^8+...
\cr\cr <\sigma _{2,1} \sigma _{5,1}>_5 & = &
1-2z^3-6z^4-10z^5-28z^6-50z^7-118z^8-192z^9-362z^{10}+... \cr\cr
<\sigma _{2,1} \sigma _{6,1}>_6 & = &
1-2z^3-6z^4-10z^5-28z^6-50z^7-118z^8-194z^9-380z^{10} \cr\cr & &
-540z^{11}-928z^{12}+... \eeqs
These series expansions approach the expected result for ${\cal M}
\to \infty$ which is
\beqs \lim _{{\cal M} \to \infty} <\sigma _{2,1} \sigma _{{\cal
M},1}> & = &
1-2z^3-6z^4-10z^5-28z^6-50z^7-118z^8-194z^9-380z^{10}-542z^{11}
\cr\cr & & -950z^{12}-1142z^{13}-1960z^{14}+... \eeqs
Notice that the coefficient of $z^3$ term is half the value of the
coefficient of $z^3$ term of the boundary-boundary case.

In general, the matrix for $<\sigma _{m^\prime+1,1} \sigma _{{\cal
M}-m,1}>$ corresponds to the matrix for $<\sigma _{1,1} \sigma
_{{\cal M},1}>$ with the first $m^\prime$ columns and rows and the
last $m$ columns and rows removed, and we expect that $\lim
_{{\cal M} \to \infty} <\sigma _{m^\prime+1,1} \sigma _{{\cal
M}-m,1}>$ is the same as the product of the $(m^\prime +1)$-th row
magnetization and the $(m+1)$-th row magnetization.

As another example, the matrix for the boundary-third row
correlation function is the submatrix of the matrix for $<\sigma
_{1,1} \sigma _{{\cal M},1}>$ with the first two columns and rows
removed. The first three series expansions of $<\sigma _{3,1}
\sigma _{{\cal M},1}>_{\cal M}$ with finite ${\cal M}$ are given
by
\beqs <\sigma _{3,1} \sigma _{4,1}>_4 & = &
1-2z^3-6z^4-6z^5-12z^6-4z^7-16z^8+... \cr\cr <\sigma _{3,1} \sigma
_{5,1}>_5 & = &
1-2z^3-6z^4-8z^5-22z^6-22z^7-70z^8-84z^9-260z^{10}+... \cr\cr
<\sigma _{3,1} \sigma _{6,1}>_6 & = &
1-2z^3-6z^4-8z^5-22z^6-24z^7-84z^8-122z^9-420z^{10}-712z^{11}
\cr\cr & & -1774z^{12}+... \eeqs
In this case, the coefficients up to $z^6$ terms are the same as
the corresponding coefficients for $M_{\scriptsize
\mbox{s,1}}M_{\scriptsize \mbox{s,bulk}}$ given in eqn.
(\ref{bms}). We expect that the number of such leading terms in
the series expansion of $\lim _{{\cal M} \to \infty} <\sigma
_{m+1,1} \sigma _{{\cal M},1}>$ as agree with the corresponding
terms for $M_{\scriptsize \mbox{s,1}}M_{\scriptsize
\mbox{s,bulk}}$ increases monotonically as $m$ increases.

\subsection{Boundary-middle case}

For the correlation function between one spin on a free boundary
and the other spin on the middle row, substitute $m^\prime = 1$
and $m =[{\cal M}/2]$ in eqn. (\ref{generalcf}). The matrix for
$<\sigma _{1,1} \sigma _{[{\cal M}/2],1}>$ corresponds to the
upper-left quarter of the matrix for $<\sigma _{1,1} \sigma
_{{\cal M},1}>$. In the limit ${\cal M} \to \infty$, the two
factors of eqns. (\ref{elementll}) and (\ref{elementur})
containing $\alpha$ to negative infinite powers are equal to 1. As
before, $\lim _{{\cal M} \to \infty} <\sigma _{1,1} \sigma
_{[{\cal M}/2],1}>_{\ell}$ denotes the determinant of the $\ell
\times \ell$ submatrix of the matrix for $\lim _{{\cal M} \to
\infty} <\sigma _{1,1} \sigma _{[{\cal M}/2],1}>$ with first
$\ell$ columns and first $\ell$ rows. The expansion formulas for
small $\ell$ are given as
\beqs \lim _{{\cal M} \to \infty} <\sigma _{1,1} \sigma _{[{\cal
M}/2],1}>_1 & = &
1-2z^3-6z^4-6z^5-12z^6-2z^7-6z^8+34z^9+36z^{10}+162z^{11} \cr\cr &
& +170z^{12}+... \cr\cr \lim _{{\cal M} \to \infty} <\sigma _{1,1}
\sigma _{[{\cal M}/2],1}>_2 & = &
1-2z^3-6z^4-8z^5-22z^6-20z^7-60z^8-20z^9-94z^{10}+168z^{11} \cr\cr
& & +114z^{12}+... \cr\cr \lim _{{\cal M} \to \infty} <\sigma
_{1,1} \sigma _{[{\cal M}/2],1}>_3 & = &
1-2z^3-6z^4-8z^5-22z^6-22z^7-74z^8-58z^9-254z^{10}-110z^{11}
\cr\cr & & -728z^{12}+... \cr\cr \lim _{{\cal M} \to \infty}
<\sigma _{1,1} \sigma _{[{\cal M}/2],1}>_4 & = &
1-2z^3-6z^4-8z^5-22z^6-22z^7-74z^8-60z^9-272z^{10}-176z^{11}
\cr\cr & & -1090z^{12}+... \cr\cr \lim _{{\cal M} \to \infty}
<\sigma _{1,1} \sigma _{[{\cal M}/2],1}>_5 & = &
1-2z^3-6z^4-8z^5-22z^6-22z^7-74z^8-60z^9-272z^{10}-178z^{11}
\cr\cr & & -1112z^{12}+... \eeqs
which should be compared with the exact result,
\beqs \lim _{{\cal M} \to \infty} <\sigma _{1,1} \sigma _{[{\cal
M}/2],1}> & = & \sqrt{\frac{(1 - 2z-z^2)(1+z^2)}{(1-z)^3(1+z)}}
\left [1 - \frac{16z^4}{(1-z^2)^4} \right ]^{1/8} \cr\cr & = &
1-2z^3-6z^4-8z^5-22z^6-22z^7-74z^8-60z^9-272z^{10}-178z^{11}
\cr\cr & & -1112z^{12}-580z^{13}-4918z^{14}-2018z^{15}+...
\label{bms} \eeqs
By D log Pad{\' e} and with $z_c = \sqrt{2}-1$, one only needs
terms up to $z^{12}$ to obtain the exact result because
\beq \frac{d}{dz} \ln \Bigl ( \lim _{{\cal M} \to \infty} <\sigma
_{1,1} \sigma _{[{\cal M}/2],1}> \Bigr ) =
\frac{-2z^2(3+12z+2z^2-z^4)}{(1-z^4)(1-2z-z^2)(1+2z-z^2)} \ . \eeq

\subsection{Second row-middle and higher row-middle cases}

For the second row-middle correlation function, $<\sigma _{2,1}
\sigma _{[{\cal M}/2],1}>$, its matrix corresponds to the matrix
for $<\sigma _{1,1} \sigma _{[{\cal M}/2],1}>$ with the first
column and row removed. Now we calculate the quantity $\lim
_{{\cal M} \to \infty} <\sigma _{2,1} \sigma _{[{\cal
M}/2],1}>_\ell$ which is defined by the determinant of the $\ell
\times \ell$ submatrix of the matrix for $\lim _{{\cal M} \to
\infty} <\sigma _{2,1} \sigma _{[{\cal M}/2],1}>$ with the first
$\ell$ columns and the first $\ell$ rows, we calculate
\beqs \lim _{{\cal M} \to \infty} <\sigma _{2,1} \sigma _{[{\cal
M}/2],1}>_1 & = &
1-4z^4-2z^5-18z^6-26z^7-74z^8-94z^9-166z^{10}-86z^{11} \cr\cr & &
-40z^{12}+... \cr\cr \lim _{{\cal M} \to \infty} <\sigma _{2,1}
\sigma _{[{\cal M}/2],1}>_2 & = &
1-4z^4-2z^5-22z^6-28z^7-108z^8-148z^9-402z^{10}-484z^{11} \cr\cr &
& -1082z^{12}+... \cr\cr \lim _{{\cal M} \to \infty} <\sigma
_{2,1} \sigma _{[{\cal M}/2],1}>_3 & = &
1-4z^4-2z^5-22z^6-28z^7-112z^8-150z^9-460z^{10}-574z^{11} \cr\cr &
& -1688z^{12}+... \cr\cr \lim _{{\cal M} \to \infty} <\sigma
_{2,1} \sigma _{[{\cal M}/2],1}>_4 & = &
1-4z^4-2z^5-22z^6-28z^7-112z^8-150z^9-464z^{10}-576z^{11} \cr\cr &
& -1778z^{12}+... \eeqs
These series expansions approach the expected result which is
\beqs \lim _{{\cal M} \to \infty} <\sigma _{2,1} \sigma _{[{\cal
M}/2],1}> & = &
1-4z^4-2z^5-22z^6-28z^7-112z^8-150z^9-464z^{10}-576z^{11} \cr\cr &
& -1782z^{12}-2014z^{13}-7054z^{14}-7240z^{15}+... \eeqs
Notice that there is no $z^3$ term in contract with the
boundary-middle case in eqn. (\ref{bms}), and the coefficient of
$z^4$ term is the same as the coefficient of $z^4$ term for the
bulk case in eqn. (\ref{bulks}).

In general, the matrix for $<\sigma _{m+1,1} \sigma _{[{\cal
M}/2],1}>$ corresponds to the matrix for $<\sigma _{1,1} \sigma
_{[{\cal M}/2],1}>$ with the first $m$ columns and rows removed,
and we expect that $\lim _{{\cal M} \to \infty} <\sigma _{m+1,1}
\sigma _{[{\cal M}/2],1}>$ is the same as the product of the bulk
magnetization and the $(m+1)$-th row magnetization.

As the next example, the matrix for the third row-middle
correlation function is the submatrix of the matrix for $\lim
_{{\cal M} \to \infty} <\sigma _{1,1} \sigma _{[{\cal M}/2],1}>$
with the first two columns and rows removed. The first three
series expansions of $\lim _{{\cal M} \to \infty} <\sigma _{3,1}
\sigma _{[{\cal M}/2],1}>_{\ell}$ are given by
\beqs \lim _{{\cal M} \to \infty} <\sigma _{3,1} \sigma _{[{\cal
M}/2],1}>_1 & = &
1-4z^4-12z^6-2z^7-46z^8-62z^9-276z^{10}-518z^{11} \cr\cr & &
-1278z^{12}+... \cr\cr \lim _{{\cal M} \to \infty} <\sigma _{3,1}
\sigma _{[{\cal M}/2],1}>_2 & = &
1-4z^4-16z^6-2z^7-70z^8-64z^9-394z^{10}-624z^{11} \cr\cr & &
-2052z^{12}+... \cr\cr \lim _{{\cal M} \to \infty} <\sigma _{3,1}
\sigma _{[{\cal M}/2],1}>_3 & = &
1-4z^4-16z^6-2z^7-74z^8-64z^9-438z^{10}-626z^{11} \cr\cr & &
-2362z^{12}+... \eeqs
In this case, no $z^3$ and $z^5$ terms appear, and the
coefficients of $z^4$ and $z^6$ terms are the same as the
corresponding coefficients for the bulk case in eqn.
(\ref{bulks}). Therefore, we expect that the number of such
leading terms in the series expansion of $\lim _{{\cal M} \to
\infty} <\sigma _{m+1,1} \sigma _{[{\cal M}/2],1}>$ as agree with
the corresponding terms for the bulk case increases again
monotonically as $m$ increases.

\section{Conclusions}

In conclusion, long-range correlation functions between two spins
on the same column of the rectangular Ising model with cyclic
boundary conditions were studied. The low-temperature series
expansions of the correlation functions were presented when the
spin-spin couplings are the same in both directions. We confirmed
that as the two spins are infinitely far from each other, the
correlation function is equal to the product of the row
magnetizations of the corresponding spins. The approach of this
$m$-th row correlation function to the bulk correlation function
for increasing $m$ could be understood using low-temperature
series expansions. Namely the dominant terms of their series
expansions are the same and the number of these terms increases
monotonically as $m$ increases.

\vspace*{1cm}

Acknowledgment: One of the authors (S.-C. C.) would like to thank
the support from Nishina Memorial Foundation and Inoue Foundation.

\vfill \eject
\end{document}